\documentclass[jkps,preprint,fleqn,showpacs,showkeys]{revtex4}
\usepackage[pdftex]{graphicx}
\usepackage{amssymb}
\usepackage{amsmath}
\usepackage{bm}
\usepackage[font=normalsize]{caption}
\begin{document}
\setcounter{page}{1}

\title{Quark Gluon Jet Discrimination with Weakly Supervised Learning}

\author{Jason Sang Hun \surname{Lee}}
\email{jason.lee@uos.ac.kr}
\affiliation{Department of Physics, University of Seoul, Seoul 02504}

\author{Sang Man \surname{Lee}}
\affiliation{Department of Physics, University of Seoul, Seoul 02504}

\author{Yunjae \surname{Lee}}
\email{yunjae.lee@cern.ch}
\affiliation{Department of Physics, University of Seoul, Seoul 02504}

\author{Inkyu \surname{Park}}
\affiliation{Department of Physics, University of Seoul, Seoul 02504}

\author{Ian James \surname{Watson}}
\email{ian.james.watson@cern.ch}
\affiliation{Department of Physics, University of Seoul, Seoul 02504}

\author{Seungjin \surname{Yang}}
\affiliation{Department of Physics, University of Seoul, Seoul 02504}

\begin{abstract}
  Deep learning techniques are currently being investigated for high energy physics experiments, to tackle a wide range of problems, with quark and gluon discrimination becoming a benchmark for new algorithms.
  One weakness is the traditional reliance on Monte Carlo simulations, which may not be well modelled at the detail required by deep learning algorithms.
  The weakly supervised learning paradigm gives an alternate route to classification, by using samples with different quark--gluon proportions instead of fully labeled samples.
  The paradigm has, therefore, huge potential for particle physics classification problems as these weakly supervised learning methods can be applied directly to collision data.
  In this study, we show that realistically simulated samples of dijet and Z+jet events can be used to discriminate between quark and gluon jets by using weakly supervised learning. 
  We implement and compare the performance of weakly supervised learning for quark--gluon jet classification using three different machine learning methods: the jet image-based convolutional neural network, the particle-based recurrent neural network and and the feature-based boosted decision tree.
\end{abstract}



\maketitle

\section{Introduction}

The use of machine learning techniques in high energy physics has been of major interest in recent years due to its potential to improve the analysis of particle collision data.
One specific area in which machine learning is used for improvement is the discrimination between quark-initiated and gluon-initiated jets \cite{Komiske:2016rsd,guest2016jet}.
Though these machine learning techniques show excellent performance, they heavily rely on Monte Carlo (MC) simulations for input, as they are trained on the microscopic details of the simulation, which may not be well-modelled due to the non-perturbative nature of Quantum Chromodynamics (QCD) at low energies.
Thus, the performance of these methods can be sub-optimal when applied to real data, and care is needed.

In contrast, weakly supervised paradigms, such as Classification Without Labels (CWoLa) \cite{metodiev2017classification} and Learning from Label Proportions (LLP) \cite{quadrianto2009estimating,patrini2014almost}, can alleviate these issues as they can be used as data-driven classifiers.
This is done as they allow training using samples that are mixtures of quark and gluon events of different proportions rather than requiring pure, labeled quark and gluon samples.
In the CWoLa method, you train a classifier to distinguish between quark-enriched and gluon-enriched samples.
Under the condition that the only difference between the two samples are the quark--gluon proportions (and not the features of the quark or gluon in each sample), training a classifier to distinguish the two samples is equivalent to training a classifier to distinguish a quark and from a gluon.
The CWoLa technique is beginning to be used in LHC analyses; for example in the CMS \(t\bar{t}b\bar{b}\) analysis, it has been used to distinguish the multijet background, which is difficult to model because of the high number of jets \cite{CMS-PAS-TOP-18-011}.

The CWoLa method is simple to apply to any machine learning algorithm as it allows training without any truth information, such as quark--gluon labels or the class proportions of the mixed samples, but uses the same techniques as for standard machine learning with labels.
This allows for direct use with dijet and Z+jet samples for quark--gluon jets as they have different quark and gluon fractions \cite{gras2017systematics,gallicchio2011pure}.
A classifier that is trained to distinguish between two mixed samples, which could be made directly from collision data even though we use simulations for this study, is also able to optimally discriminate between the quark jet and the gluon jet processes, in the limit where the only difference between the samples is the quark-gluon jet fraction \cite{metodiev2017classification,komiske2018learning}.

Three machine learning methods, Convolutional Neural Network (CNN), Recurrent Neural Network (RNN) and Boosted Decision Tree (BDT) are used for weakly supervised learning.
The CNN, which is used as an image analysis and classification technique, is able to operate on jet images, which are built from energy flow information, by using successive layers of convolution filters to distinguish between gluon jets and quark jets \cite{Cogan:2014oua, de2016jet, Komiske:2016rsd}.
On the other hand, the RNN, which is used as a language or sound analysis technique, is used to operate on ordered sequences of particle information vectors from jets \cite{guest2016jet} with recurrent layers.
Finally, BDTs are a traditionally used particle physics machine learning approach that operates on high-level features extracted from the jets.

This study explores the possible use of the CWoLa method in high energy particle physics (HEP) by applying the paradigm to the quark--gluon jet-discrimination problem and comparing the results obtained to those obtained using the fully supervised paradigms.
We extend previous weakly supervised learning studies by using realistic samples of dijet and Z+jet events, rather than simply creating mixed quark and gluon jet samples from a single source.
We also test the performance by using not just previously investigated BDT classifiers but also with deep learning CNN and RNN classifiers, as any machine learning algorithm is compatible with the CWoLa method.

\section{Monte Carlo Models}

MadGraph5\_aMCatNLO v2.6.3.2 \cite{Alwall:2011uj} was used to generate the hard process for each sample. 
These hard processes were then passed to PYTHIA 8.240 for the simulation of parton showering and underlying event generation \cite{Sjostrand:2014zea}.
We used the default PYTHIA settings, which correspond to the Monash 2013 tune \cite{monash2013}.
Finally, the fast detector simulator DELPHES 3.4.1 \cite{deFavereau:2013fsa} was used to approximate the response of the CMS particle-flow reconstruction algorithm \cite{cms2009particle}.
For the jet clustering, DELPHES uses the FASTJET 3.3.2 \cite{Cacciari:2011ma} package, and jets are clustered with the anti-$\it{k_{T}}$ algorithm \cite{Cacciari:2008gp} with a jet radius R = 0.4.
We used the settings from the default CMS card distributed with DELPHES, except to reduce the jet radius to 0.4, to match the current CMS value.

We generated Z+jet and dijet events by using the matrix elements indicated by Table~\ref{tab:dataset}. 
As the Z boson will be reconstructed in the dimuon channel, the Z boson is forced to decay to \(\mu^+\mu^-\).
Separate samples were produced for a realistic sample and pure samples. In the realistic sample, the outgoing partons, indicated by \(j\), were unrestricted between gluon and all light quark flavors (u, d, s). 
For the pure samples, the only parton was only allowed to be from the light quarks for the \(q\) samples or from a gluon for the \(g\) samples.
By producing dedicated pure $q\bar{q}$, $gg$ and $Zq$, $Zg$ samples, we obviate the need to assign labels based on parton matching as the jets selected from the $q\bar{q}$ and the $Zq$ samples are considered to be quark-initiated, and the jets from $gg$ and $Zg$ to be gluon-initiated.
Such an approach has been used in previous studies \cite{Komiske:2016rsd} and $\Delta R < 0.4$ matching the parton and jets is found to select more than 95\% of the sample thereby produced.
We produced samples for various \(p_T\) ranges to compare the performance on various energy scales.
That is, when producing the samples, we applied a hard cut to the parton \(p_T\) at the generation level. 
The ranges we used are 100--110 GeV, 200--220 GeV, 500--550 GeV and 1000--1100 GeV.



\begin{table}[ht]
\begin{tabular}{|c|c|c|c|c|c|}
\hline
Dataset Name & Quark-like & Gluon-like & Training & Validation & Test\\ \hline \hline
Realistic      & \(p p \rightarrow Z j\)  & $p p \rightarrow j j$ & 140K & 60K & -     \\ \hline
Pure Dijet  & $p p \rightarrow q q$  & $p p \rightarrow g g$ & 100K  & 40K & 60K     \\ \hline
Pure Z+jet  & $p p \rightarrow Z q$ & $p p \rightarrow Z g$ & 100K & 40K & 60K     \\ \hline
\end{tabular}
\caption{List of MC simulated samples used for the training, validation and test of the machine learning models. The abbreviation \(q\) stands for any light quark (\(u\), \(d\), \(s\)), \(g\) for gluon, \(j\) for any light quark or gluon, and \(p\) for any quark or gluon.  } 
\label{tab:dataset}
\end{table}

\section{Event Selection}

The event selection process closely mirrors the event selections that are used for the CMS 13 TeV quark--gluon BDT discrimination \cite{CMS-PAS-JME-16-003}.
The dijet events require the presence of at least two jets that have a transverse momentum $p_T$ greater than 30 GeV and are balanced. 
This means that events with an additional reconstructed jet are kept if the $p_T$ of the third highest $p_T$ jet is less than 30\% of the average $p_T$ of the two leading jets.
These jets must be found within an pseudorapidity ($\eta$) range \(|\eta| < 2.4\) and the azimuthal angle between the two leading jets \(\Delta\phi\) must satisfy \(\Delta\phi > 2.5\). 

For Z+jet events, the jet must be balanced with the Z boson without any additional high-energy activity. 
Two oppositely charged muons with $p_T$ greater than 20 GeV must be present in the event for the reconstruction of the Z boson, and combined invariant mass must be within 20 GeV of the known Z boson mass \cite{pdg}.
At least one jet is required to have a $p_T$ of at least 30 GeV, and any other jets that are reconstructed in the event must individually have less than 30\% of the $p_T$ of the dimuon system.
Finally, the leading jet and the Z boson are required to be back-to-back, and the azimuthal angle between the reconstructed Z boson and the leading jet must satisfy \(\Delta\phi > 2.1\). 

Both jets that pass the event criteria in the dijet samples are used for this study while only the leading jet that is balanced against the Z boson is used for the study on the Z+jet samples.
As the $p_T$ ranges are defined in terms of the parton $p_T$, which can differ from the resulting jet $p_T$, we remove the lowest and the highest 15\% of the jets (in terms of $p_T$) from the sample after event selection so that the jet $p_T$ ranges match the parton $p_T$ better.
Information on the daughter particles of each jets was used as an input to the machine learning models.

After the event selection, the samples were divided into training, validation and test sample sets for machine learning.
The training set is passed through the model during training, and the model learns features from the training set.
After training, we evaluate the model with the validation set to check for overtraining and to pick the best performing model.
The test set is used for extracting the actual performance of the model for an independent dataset.
To facilitate comparison, we always do the testing using pure quark and gluon samples, even for models trained using realistic samples.
That is, the model selection and overtraining check of the realistic sample is done by comparing the performance to discriminate a dijet sample from a Z+jet sample, while the testing is done by comparing the performance to discriminate a pure quark sample from a pure gluon sample.

\begin{figure}[t]
    \centering
        \includegraphics[width=.4\linewidth]{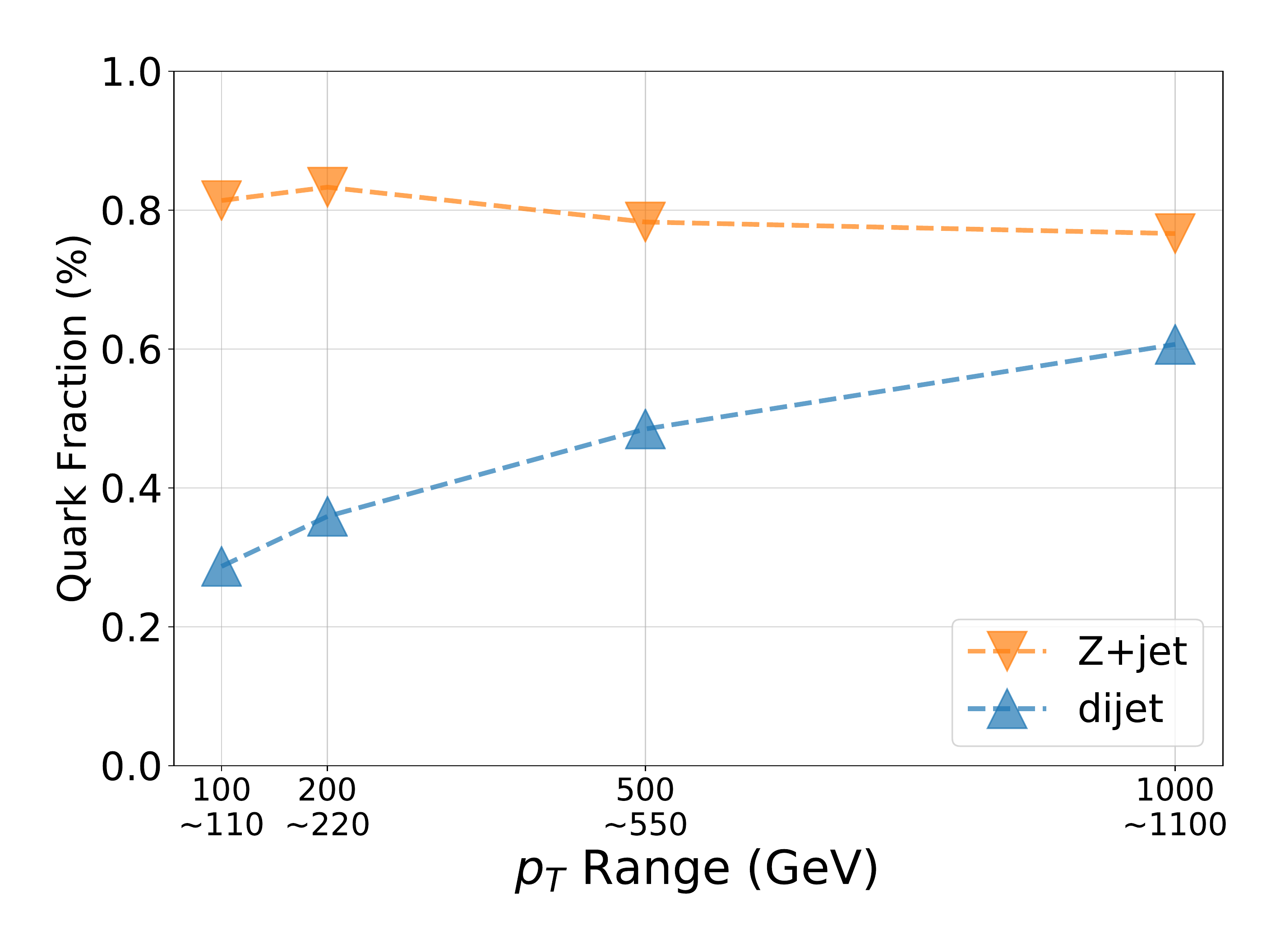}
    \caption{
    Quark fractions of the dijet and the Z+jet samples. The quark fraction of the dijet sample can be seen to increases with increasing $p_T$.}
    \label{fig:effective-cross-section}
\end{figure}

As both quark jets and gluon jets can be generated from jet production, the effective cross-section \(\mathrm{\sigma_{eff}}\), which is defined as the selection efficiency times the cross-section, is used to determine the quark jet fraction \(\mathrm{f_{q}}\) after the selection process in the realistic dijet and Z+jet samples.
The quark-jet fraction is determined by calculating
\(\mathrm{f_{q}(pp \rightarrow jj)  = \frac{\sigma_{eff}(pp \rightarrow qq) + 0.5 \cdot \sigma_{eff}(pp \rightarrow qg)}{\sigma_{eff}(pp \rightarrow jj)}}\) for dijets and
\(\mathrm{f_{q}(pp \rightarrow Z j)  = \frac{\sigma_{eff}(pp \rightarrow Zq)}{\sigma_{eff}(pp \rightarrow Z j)}}\) for Z+jet.
The quark fraction of the both the Z+jet and the dijet samples for the four different $p_T$ ranges are shown in Fig.~\ref{fig:effective-cross-section}.
The quark fractions of the dijet and the Z+jet samples have same tendencies as they do in previous studies \cite{gallicchio2011pure,gras2017systematics}. In all the $p_T$ ranges considered, the Z+jet sample is more quark-enriched than the dijet sample, with the difference decreasing with increasing $p_T$.




\begin{figure}[th]
    \centering
    \includegraphics[width=.4\linewidth]{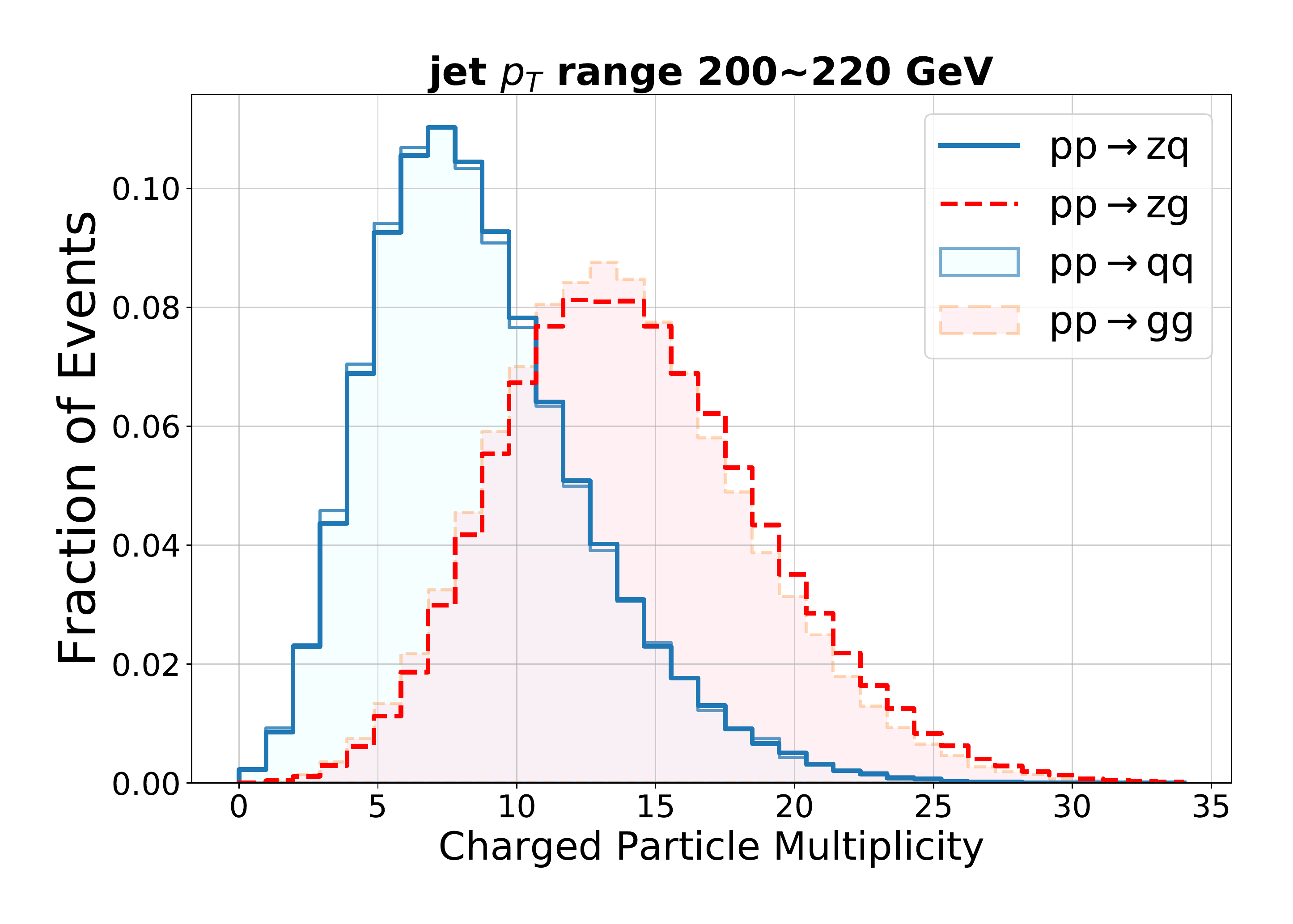}
    \includegraphics[width=.4\linewidth]{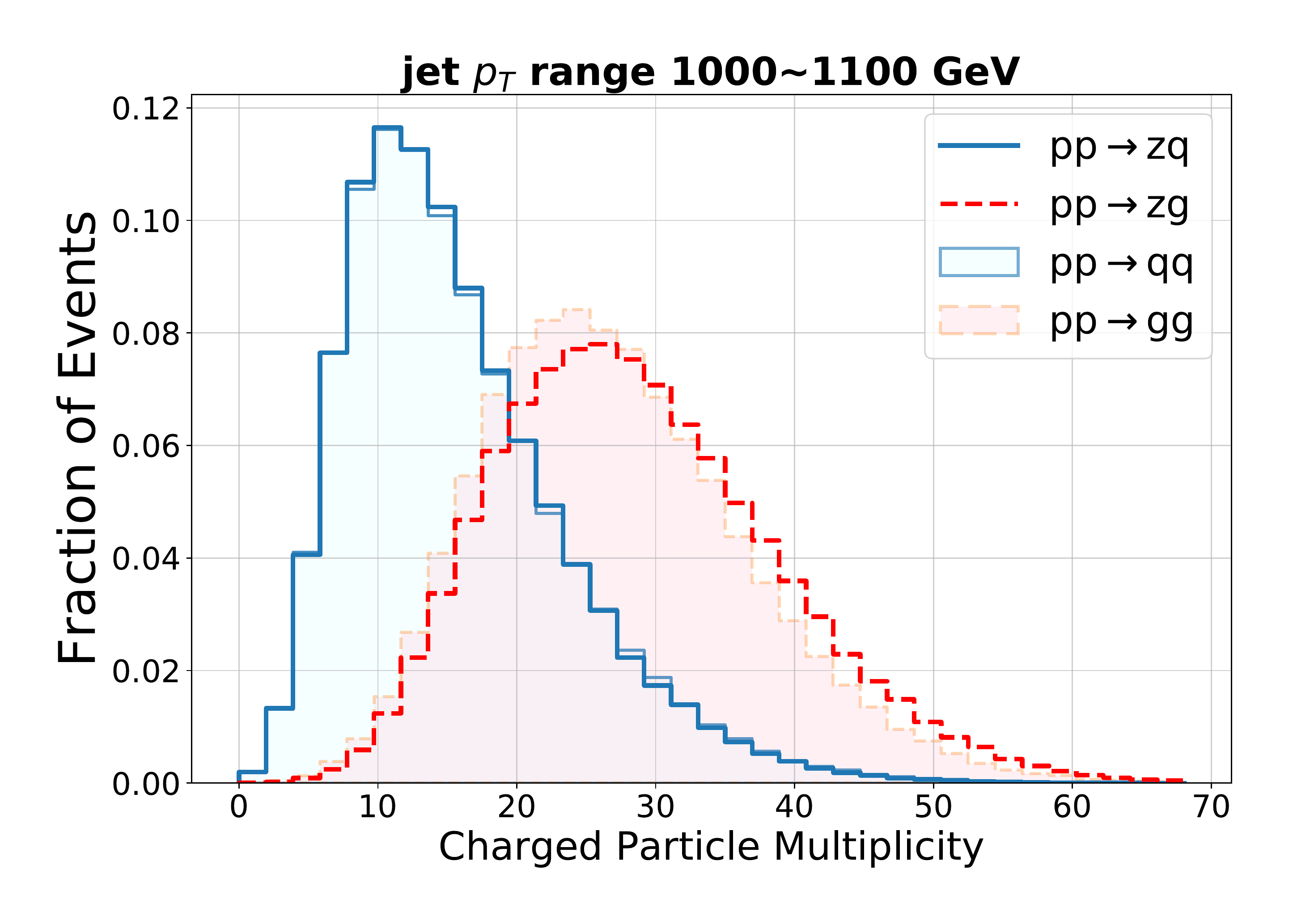}
    \includegraphics[width=.4\linewidth]{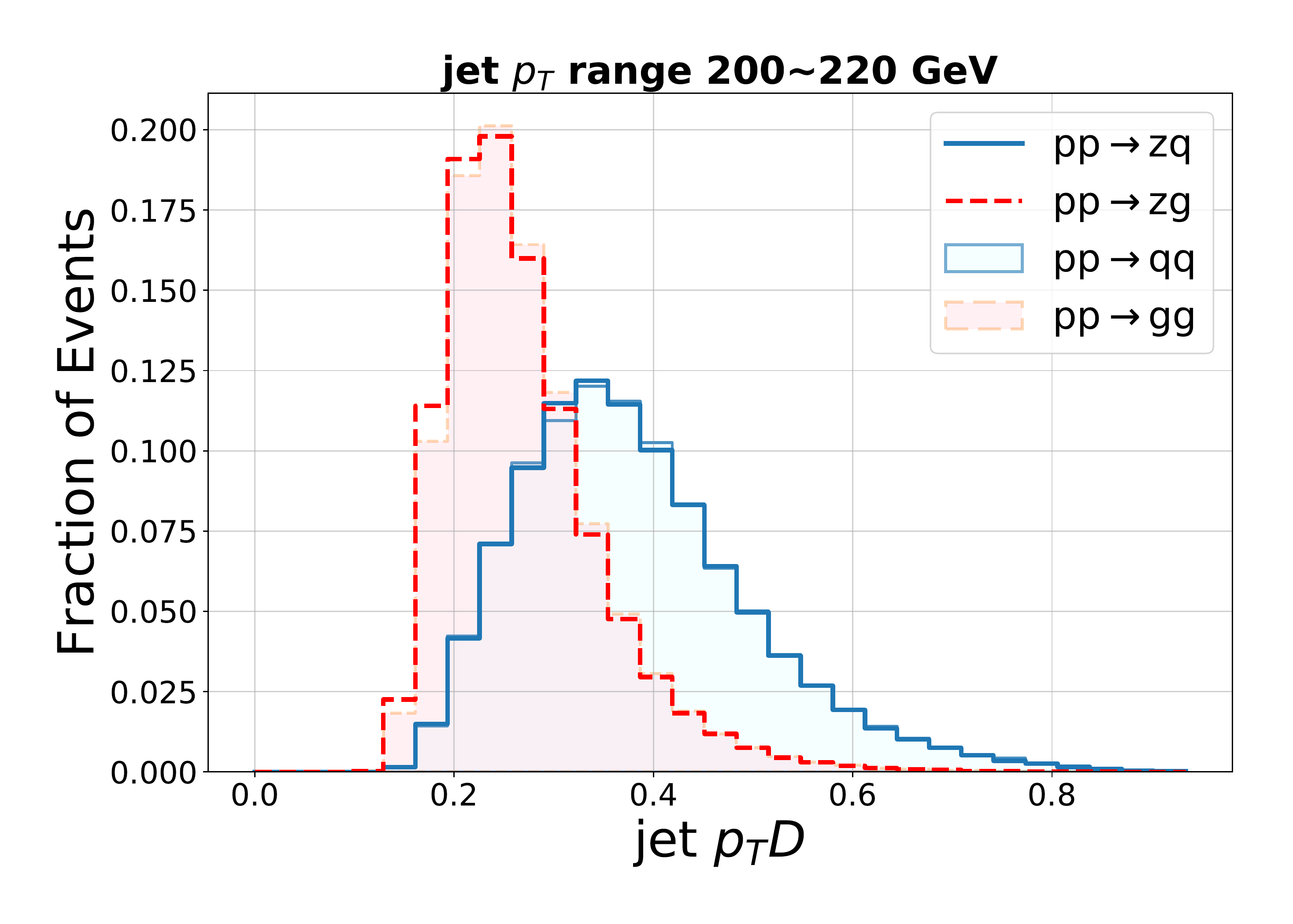}
    \includegraphics[width=.4\linewidth]{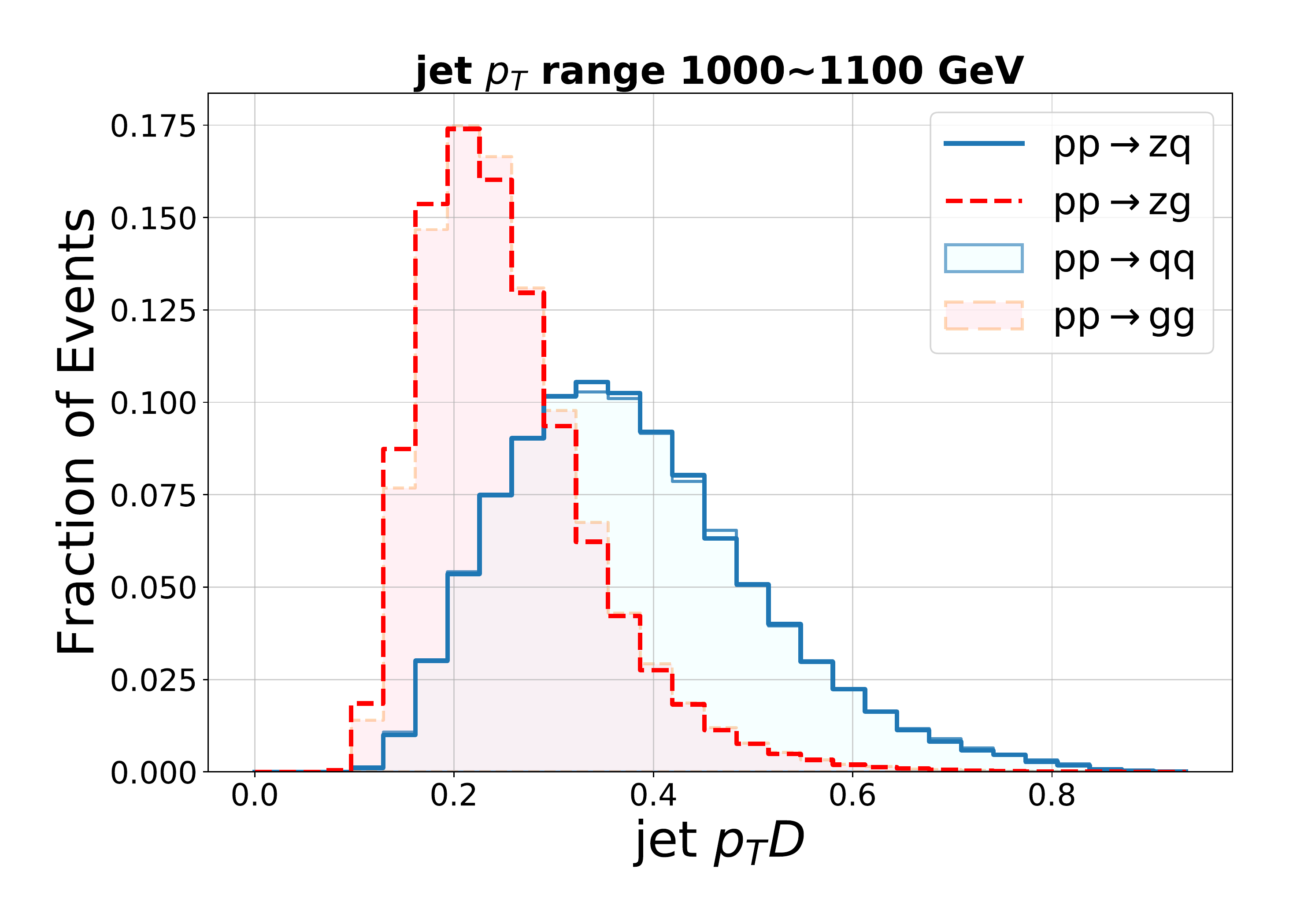}
    \caption{
      Distributions of charged particle multiplicity (top) and $p_TD$ (bottom) with jet $p_T$  in the range (left) 200--220 GeV and (right) 1000--1100 GeV. Distributions are shown for the pure quark and gluon samples in dijet and Z+jet events. 
    } 
    \label{fig:variable-dist}
\end{figure}

\section{Model description}
This section describes the three different types of data representations that are used and the corresponding machine learning algorithms. 
We use XGBoost v0.82 \cite{Chen:2016:XST:2939672.2939785} to implement Boosted Decision Trees (BDT) and Keras v2.2.4 \cite{chollet2015keras} with  TensorFlow v1.13.1 \cite{abadi2016tensorflow} backend to implement the CNN and the RNN.

\subsection{Jet Features and Boosted Decision Trees}

We trained Boosted Decision Tree (BDT) classifiers by using features selected based on the CMS study of quark--gluon jet discrimination \cite{cornelis2014quark,CMS-PAS-JME-16-003}.
We used the jet fragmentation distribution \(p_{T}D\)=$\frac{\sqrt{\sum_i p_{T,i}^2}}{\sum_i p_{T,i}}$, where $i$ is an index over the jet constituents, the charged and neutral particle multiplicities and the lengths of the major and the minor axes.
Figure ~\ref{fig:variable-dist} shows the distributions of the number of charged particles and $p_TD$ for the pure quark and gluon samples and clearly shows a separation between quarks and gluons.
The hyperparameters of the BDT were selected by a random search optimization with 5-fold cross-validation using the implementation in scikit-learn \cite{scikit-learn}.
The allowed hyperparameters (and their search ranges) are the maximum depth of the decision tree (between 1 and 5), the number of trees to fit (between 10 and 100) and the learning rate for boosting (between 0.001 to 0.5).
The default settings of XGBoost were used for the other hyperparameters.

\subsection{Jet Image and Convolutional Neural Networks}

To fully utilize the particle identification information of the reconstructed constituent particles, we built jet images of 33$\times$33 pixels in $\eta$ and $\phi$ coordinates, centered on the jet axis, from the particle-flow objects.
Ten sets of jet images have been filled (without scaling) with either the particle $p_T$ sum or the particle multiplicity for one of the particle types from charged hadrons, electrons, muons, neutral hadrons and photons \cite{lee2019quark}.
The Fully Convolutional Network (FCN) \cite{long2015fully} is used to avoid overfitting, which would usually occur due to the large number of weights that would be primarily occupied by the last fully connected layers.
Our FCN model consists of four convolutional blocks, followed by a 1x1 convolution and a softmax function.
A 5x5 convolution, followed by batch normalization, is used for the convolutional block \cite{ioffe2015batch}.
The number of output channels for the convolutional blocks are (32, 64, 64, 32). 
The network weights are initialized using Glorot uniform initialization \cite{glorot}.
Finally, the optimization is performed using root-mean-squared propagation (RMSprop) with a cross-entropy loss function.
The best model was picked by choosing the model with the best validation loss from models checkpointed at each epoch up to 20 epochs.
Averaged over the $p_T$ ranges, the best epoch was epoch 7.
After this point, based on the divergence of the validation and training loss, the model began saturating the validation and started overtraining.


\subsection{Jet Sequence and Recurrent Neural Networks}

The recurrent neural network (RNN) represents the constituents of a jet as a vector and the jet itself as a sequence of vectors.
The constituent vector is defined as an 9-tuple consisting of
$p_T$,  $\Delta \eta$, $\Delta \phi$ (with respect to the jet center), charge, and a one-hot encoding of the particle identification allowing the categories charged hadron, neutral hadron, electron, muon and photon \cite{atlas2017identification}.
To represent the constituents of the jet as a sequence, we sorted them in order of $p_{T}$, with the most energetic particle first.
As the number of constituents differs for each jet, the jet vector is padded or truncated to make all jets the same fixed length. 
For this paper, 64 constituents were used to form the sequence.
This jet sequence is fed into the Gated Recurrent Unit (GRU) \cite{cho2014learning}.
The output sequence of the GRU is fed into two subsequent fully-connected layers.
Batch normalization is used for regularization after every layer, and the stochastic gradient descent algorithm is used to minimize the cross entropy.
The best model was picked based on the validation losses from models of 60 epochs.
Averaged over the $p_T$ ranges, the best epoch was epoch 50.



\section{Results}



\begin{figure}[t]
    \centering
    \includegraphics[width=.45\textwidth]{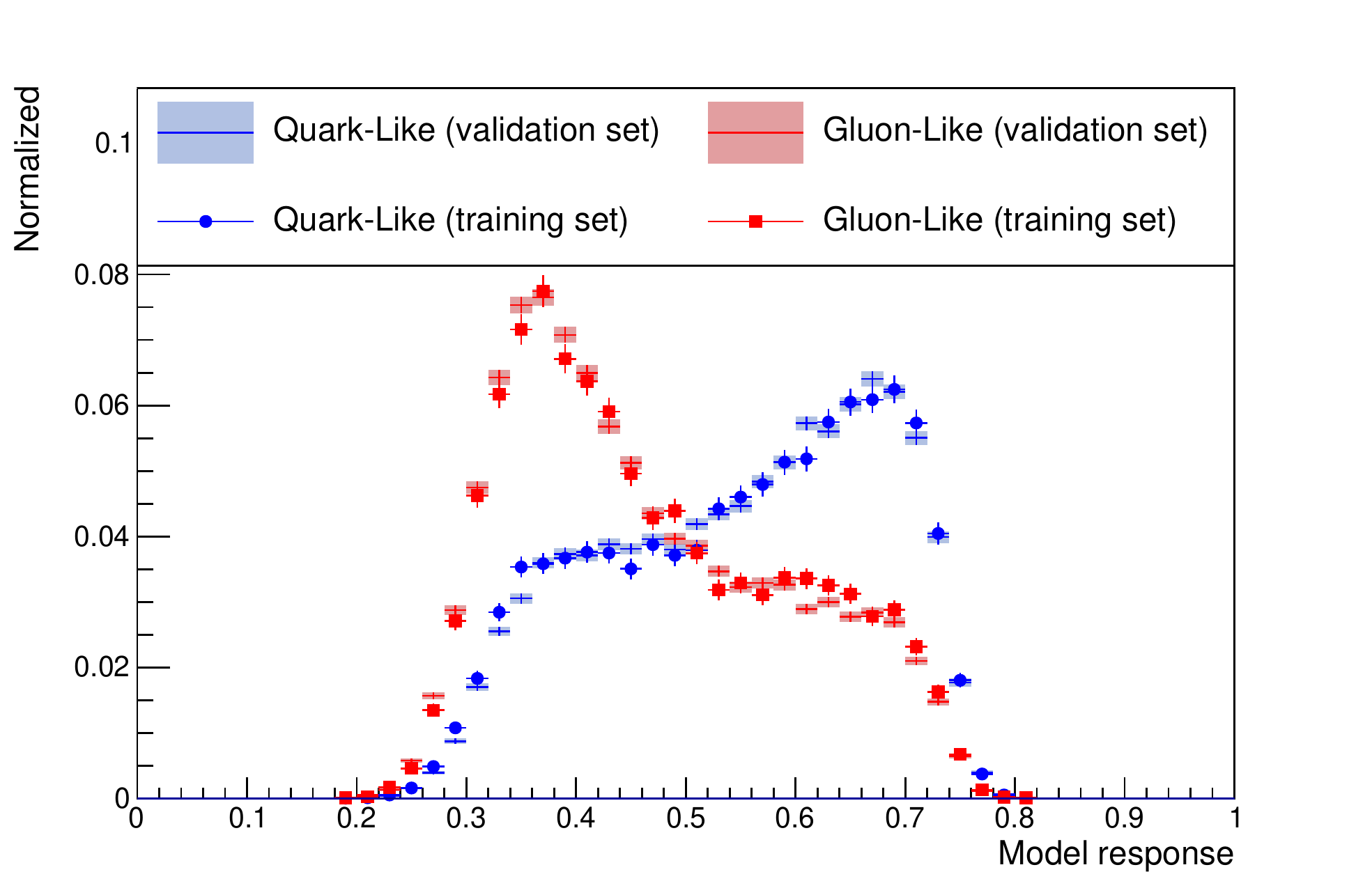}
    \includegraphics[width=.45\textwidth]{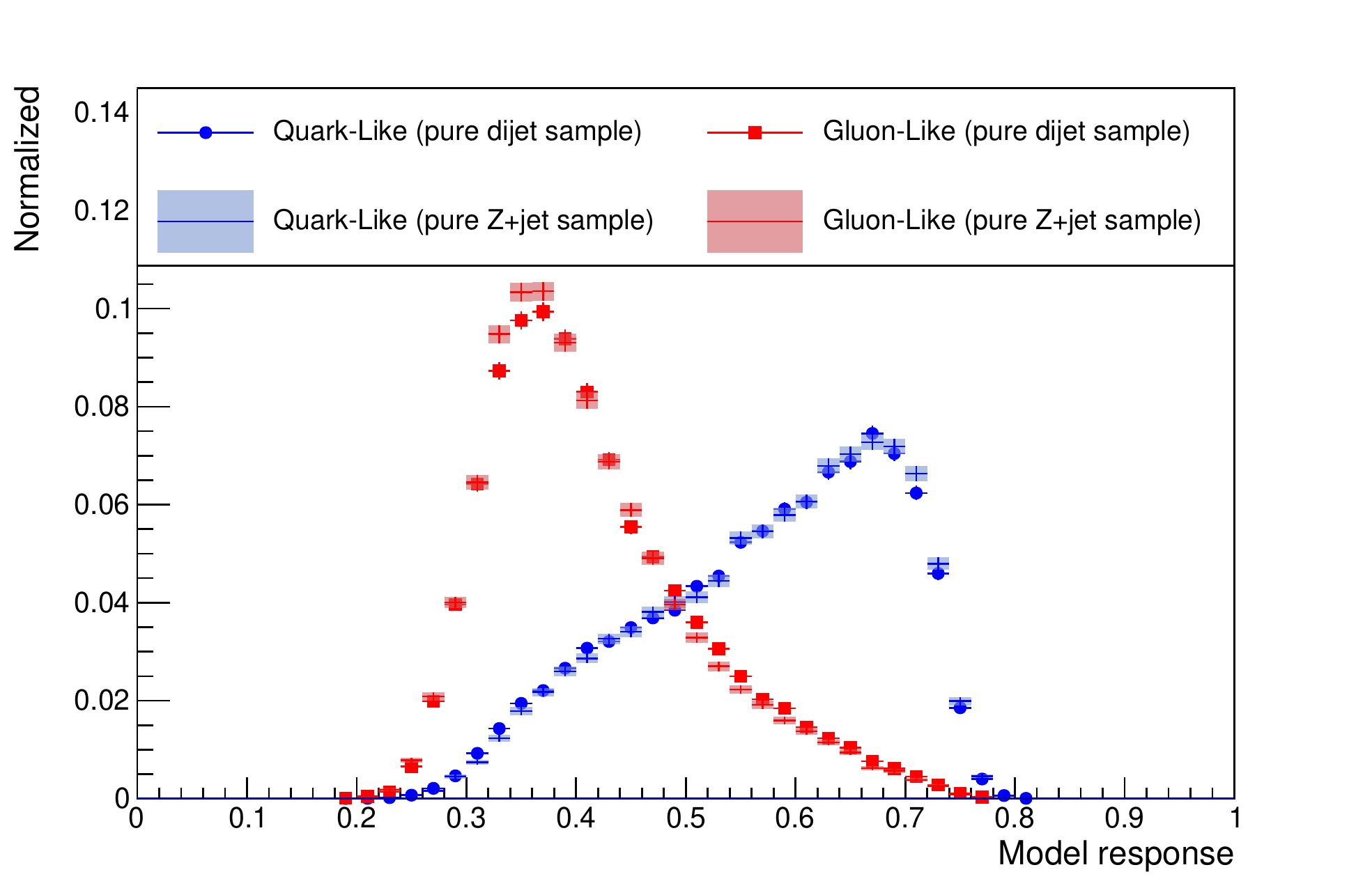}
    \caption{Weakly supervised CNN model responses to jets in the $p_T$ range of 200--220 GeV. (Left) Comparison between the training and the validation sets, which contain realistic (gluon-enriched) dijet and (quark-enriched) Z+jet samples. (Right) Comparison between the dijet and the Z+jet test sets which have pure quark and pure gluons. Clear separations between quarks and gluons can be observed in both the dijet and the Z+jet samples.}
    \label{fig:model-response}
\end{figure}

We train all the models of the previous section by using Z+jet and dijet sample discrimination in the weakly supervised paradigm as a proxy to discriminate quark and gluon jets.
That is, the models were trained to discriminate dijet from Z+jet events, which should be equivalent to training the models to discriminate between quark and gluons under the assumption that the only difference between the samples is the quark--gluon fraction.
Thus, the response should be close to zero for gluon-like jets, which have a higher fraction in the dijet sample, and close to one for quark-like jets, which have a higher fraction in the Z+jet sample.
The response of the CNN model after training is shown in Fig.~\ref{fig:model-response}, which also shows the responses to the training and the validation samples.
No significant difference is observed between the training and the validation sample responses, indicating that no overtraining occurs in this CNN model.
Also shown is the response to pure samples of quarks and gluons that have been produced separately for Z+jets and dijets and for which we see good separation.
The quark and the gluon responses are similar in both pure samples, and the realistic Z+jet and dijet responses can be clearly seen to be composed of quark and gluon jet components, as expected.

\begin{figure}[t]
\centering
\includegraphics[width=.4\linewidth]{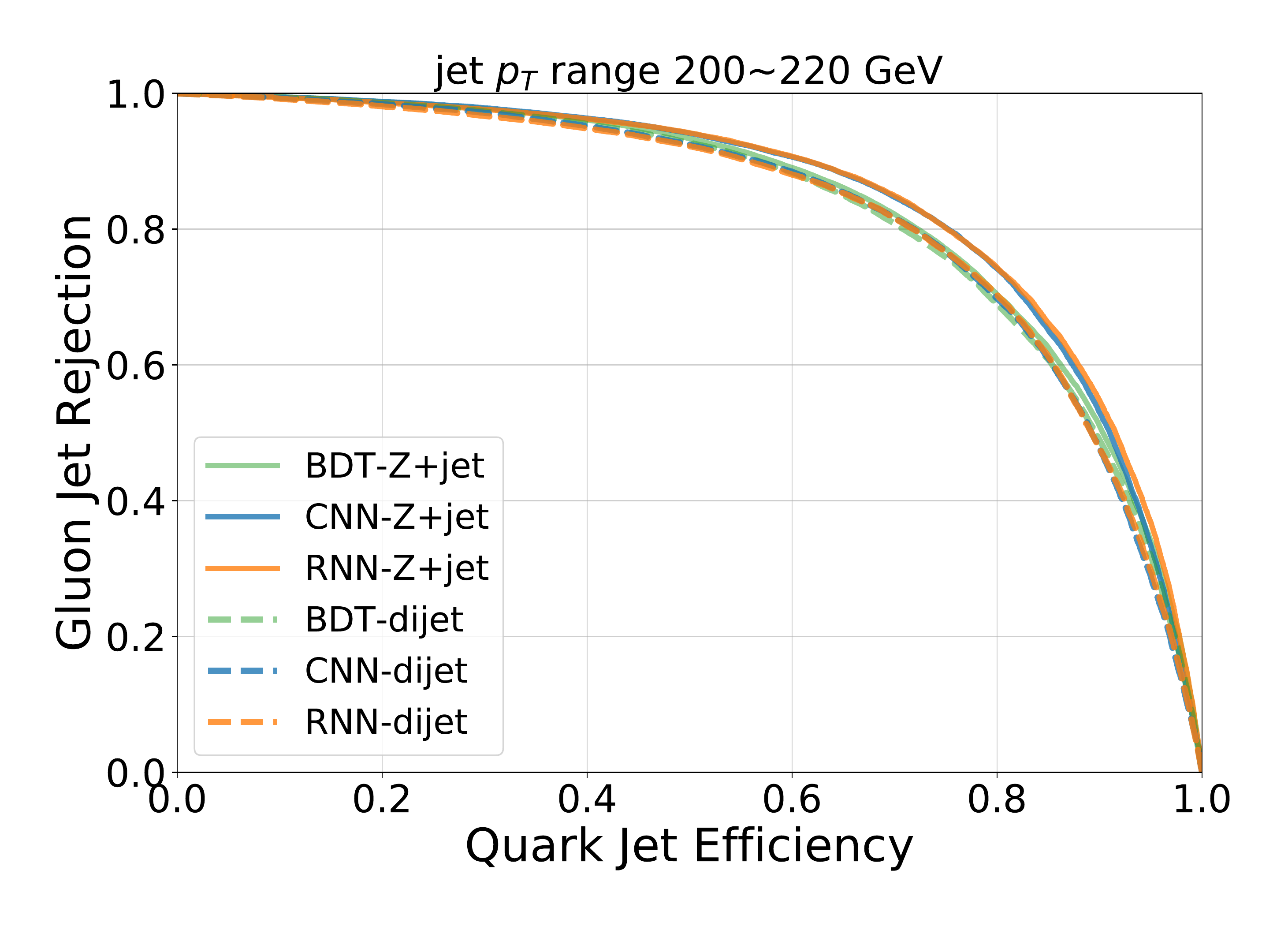}
\includegraphics[width=.4\linewidth]{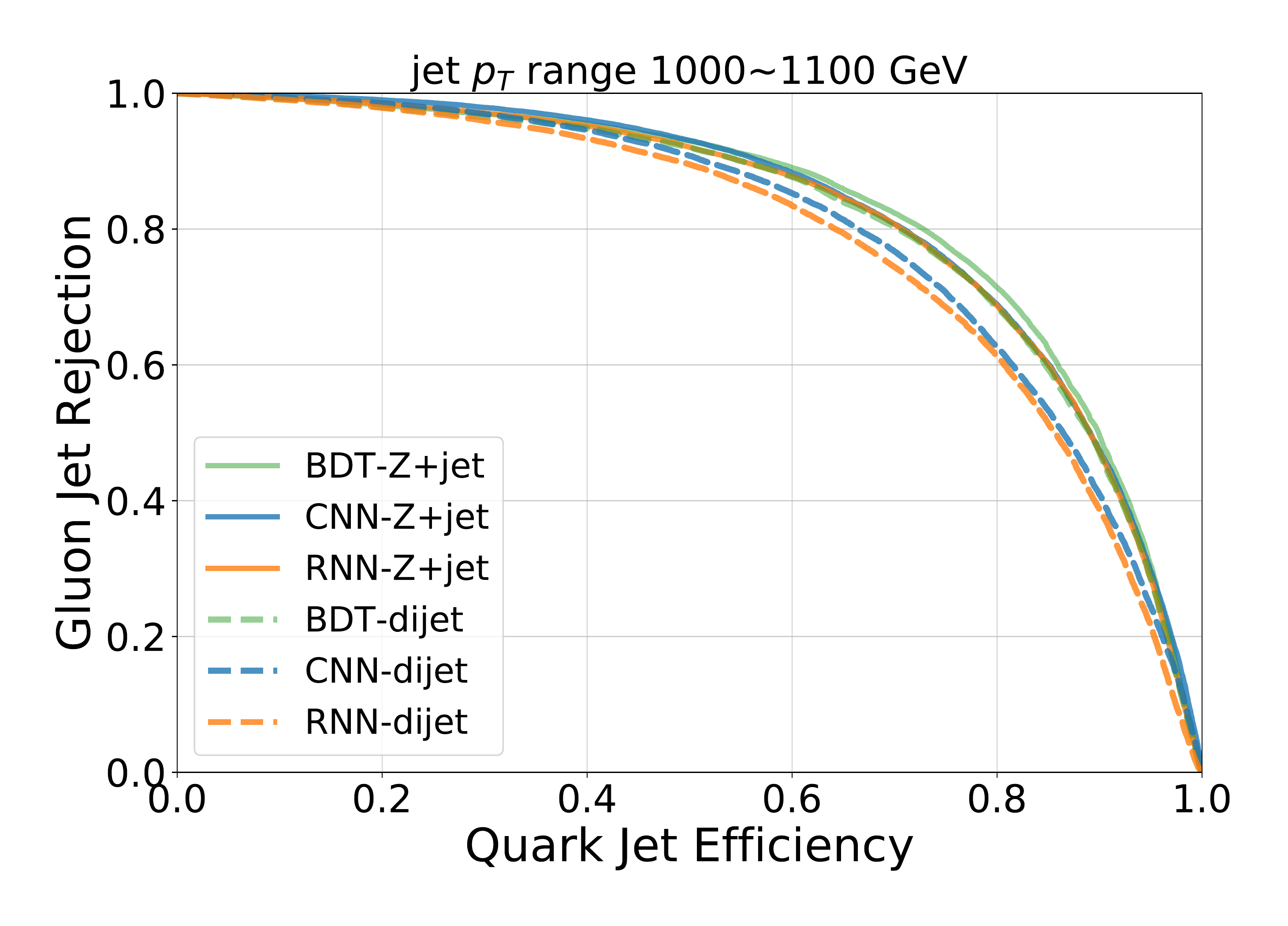}
    \caption{ROC curves for the CNN and the RNN models for the 200 GeV jet $p_T$ range (left) and the 1000 GeV jet $p_T$ range (right). Each model is trained using weakly-supervised learning with realistic samples and tested using the pure quark and the pure gluon samples.}

    \label{fig:roc}
\end{figure}

The Receiver Operation Characteristic (ROC) curves are a standard measure of the performance of a binary classifier and are created by plotting the fraction of background rejected at a given signal selection efficiency based on the classifier output.
In Fig.~\ref{fig:roc}, we compare the ROC curves from testing the efficiency of selecting from a pure quark sample versus the efficiency of rejecting from a pure gluon sample, which are produced separately in the Z+jet and the dijet channels.
For all $p_T$ ranges, the Z+jet sample shows consistently better performance than the dijet sample.
Also, the BDT shows less difference between the Z+jet and the dijet samples across the whole $p_T$ range and has better performance than the deep learning methods.
We investigated both these features of the output, as \textit{a priori}, we would expect no difference between the Z+jet and the dijet responses in the pure samples and the deep learning classifiers to perform better than the BDT, as the deep classifiers are given more information than the BDT.

\begin{figure}[t]
\centering
\includegraphics[width=.4\linewidth]{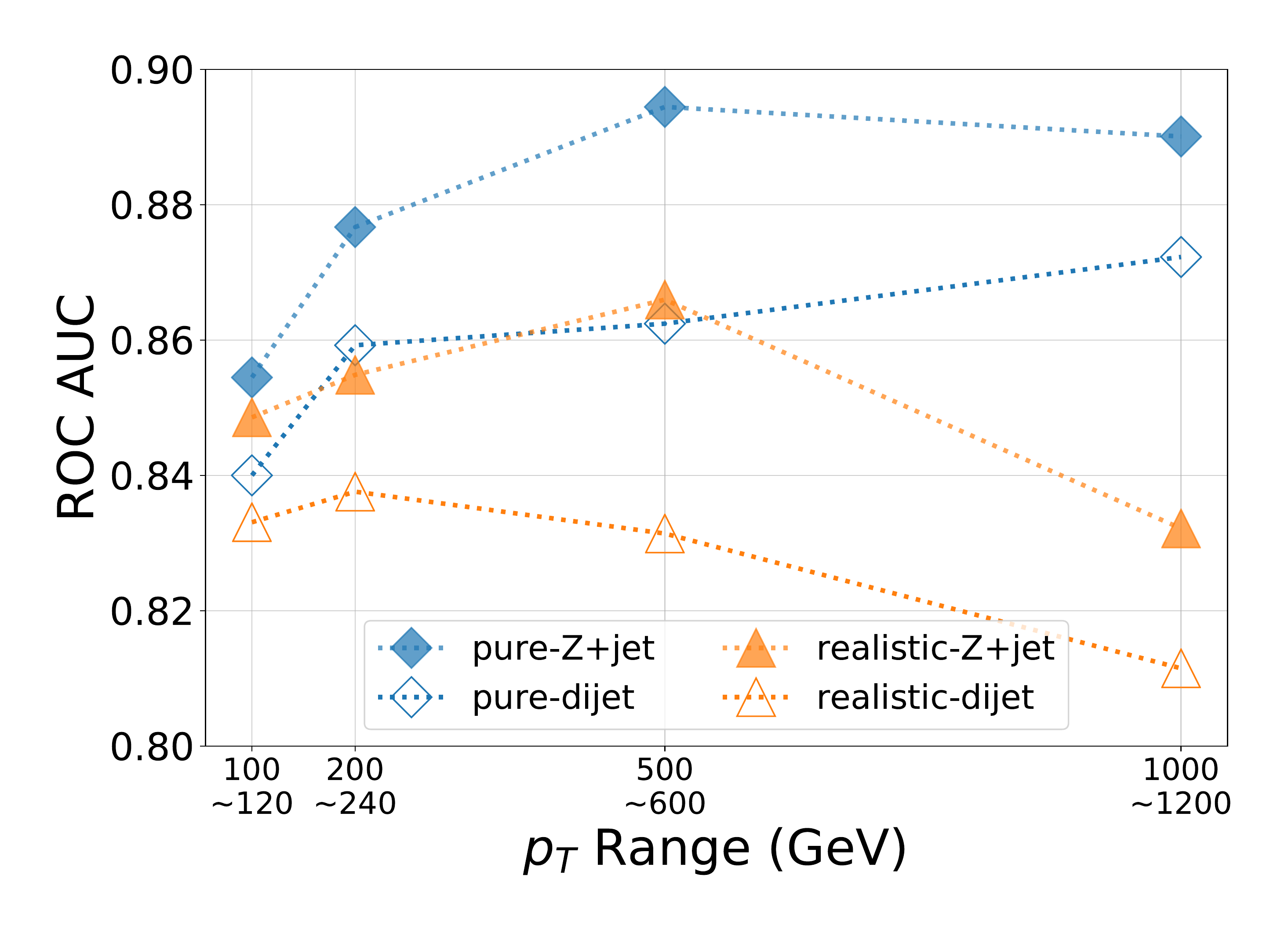}
\includegraphics[width=.4\linewidth]{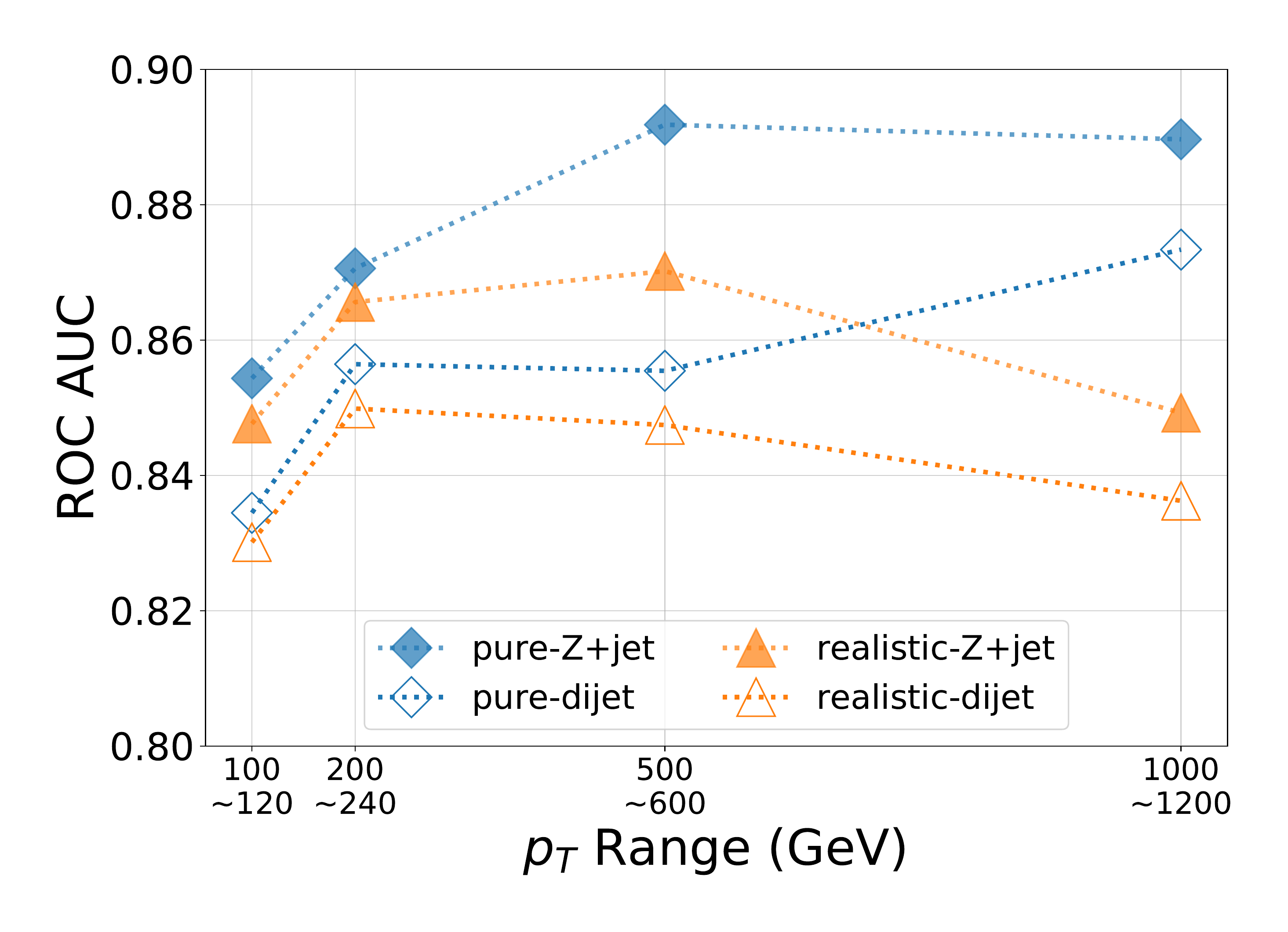}
    \caption{AUC comparisons between the fully supervised and weakly supervised neural network models. (Left) Pure Z+jet samples were used with the fully supervised CNN models while realistic Z+jet samples were used with the weakly supervised CNN models.
    (Right) Training was done on a sample with the requirement that the jet pseudorapidity satisfy $|\eta| < 1$.}
    \label{fig:performance}
\end{figure}

The Area under the ROC Curve (AUC) ranges from 0.5 for a classifier with no separation power (as much background as signal is rejected for any given efficiency level) while a perfect classifier has an AUC of one (rejecting all background for any level of signal efficiency desired).
Figure~\ref{fig:performance} compares the performances of a CNN classifier trained with fully supervised learning on the pure samples with quark and gluon labels by using the AUC.
The results are equivalent to those seen in previous quark-gluon jet studies with fully supervised training of deep learning classifiers~\cite{lee2019quark}.
The performance of the fully supervised learning is greater than that of the weakly-supervised classifiers across the whole $p_T$ range and improves with increasing $p_T$.
The decreased performance of the weakly supervised classifiers relative to that of the fully supervised sample with increasing $p_T$ is also compatible with previous results, which show that decreasing the fractional difference between the samples used for weakly-supervised training eventually leads to worse performance \cite{komiske2018learning}.
Differences in performance between the Z+jet and the dijet, equivalent to those seen in the case of the BDT with weakly supervised learning, are also seen with the fully supervised classifier.

We investigated the differences between the Z+jet and dijet quarks and gluons. Figure~\ref{fig:variable-dist} shows the distributions of a few of the BDT variables in various $p_T$ ranges.
A difference is seen between the number of charged particles produced, with more being produced in Z+jet gluons tha in dijet gluons and with the difference increasing with increasing jet $p_T$.
We created samples by changing the range of the Z and the jet $\Delta\phi$ selection, reducing the $\eta$ range of the jets and turning off the color reconnection in PYTHIA.
In all cases, the difference between the distributions remained.
Therefore, we conclude that a small physical difference exists between the gluon jets produced in the two samples, which can occur due to a color connection being allowed between the outgoing gluons in the dijet sample, but not between the gluon and the Z boson.

\begin{figure}[t]
    \centering
    \includegraphics[width=.4\linewidth]{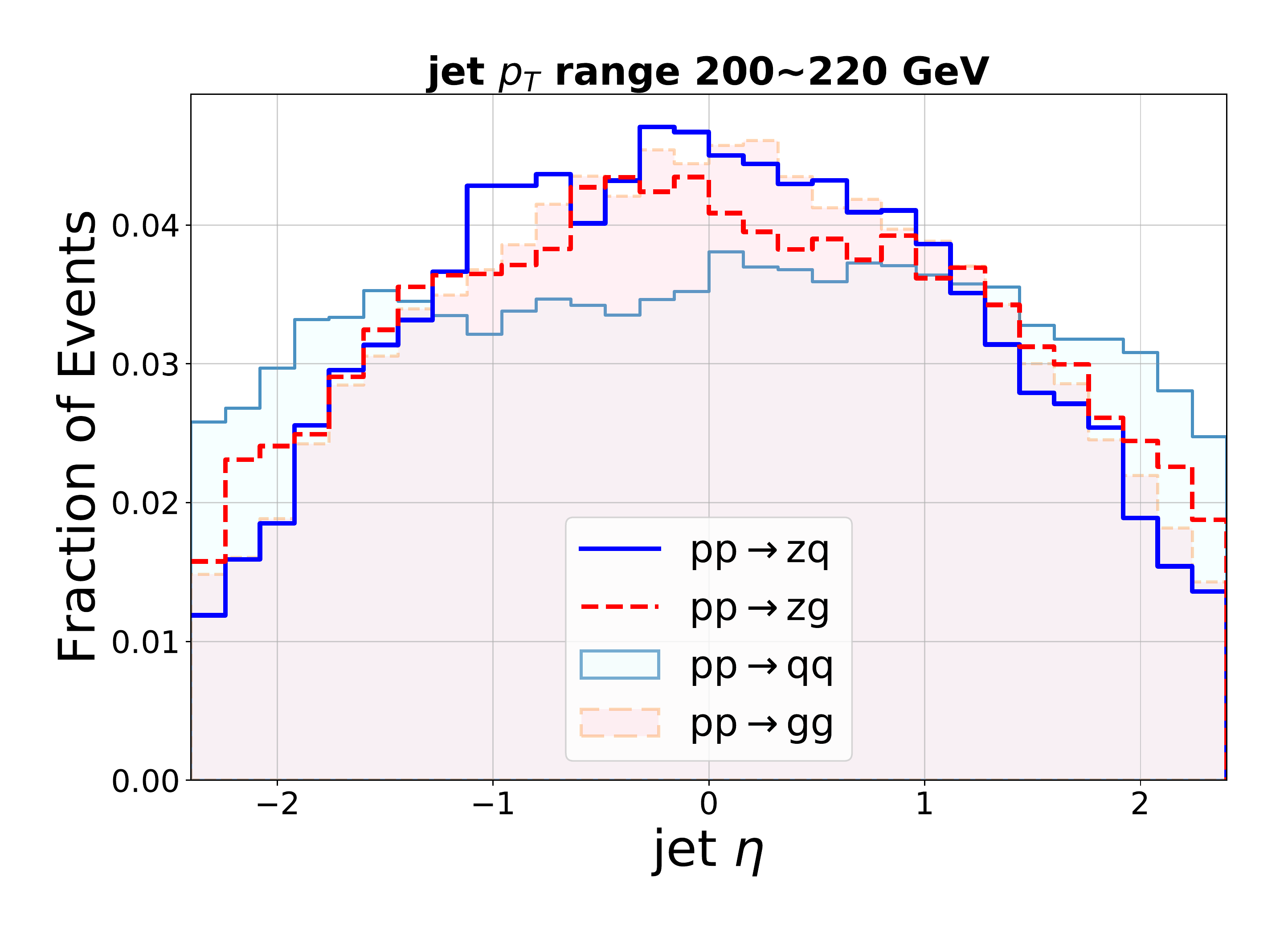}
    \includegraphics[width=.4\linewidth]{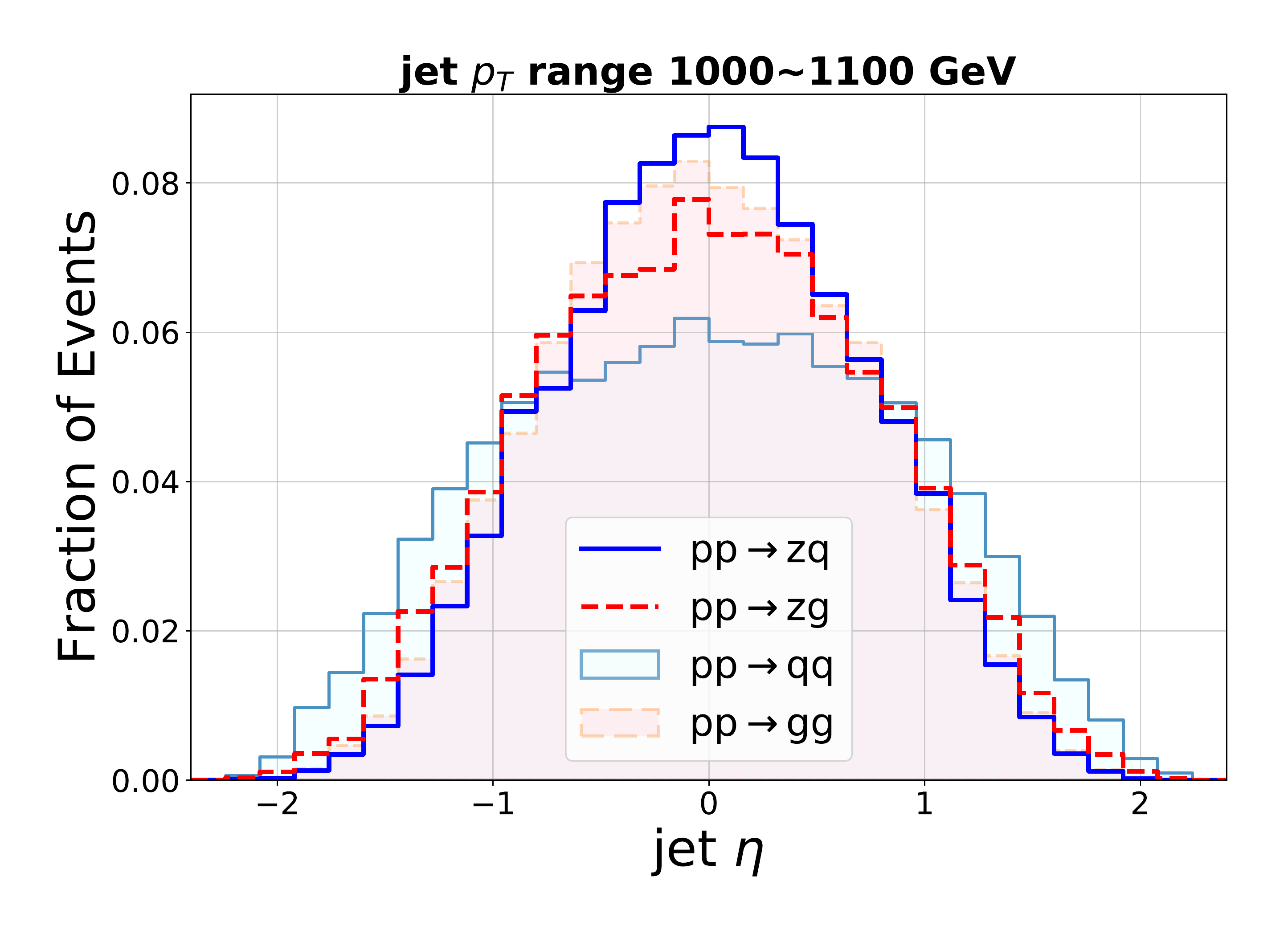}
    \caption{
        Distributions of jet $\eta$ for the $p_T$ ranges (left) 200--200 GeV and (right) 1.0--1.1 TeV.
        Quark jets from dijet events tend to distribute at larger $\eta$ than the jets in the other samples do.
      }
    \label{fig:etadist}
\end{figure}

\begin{figure}[t]
\centering
\includegraphics[width=.4\linewidth]{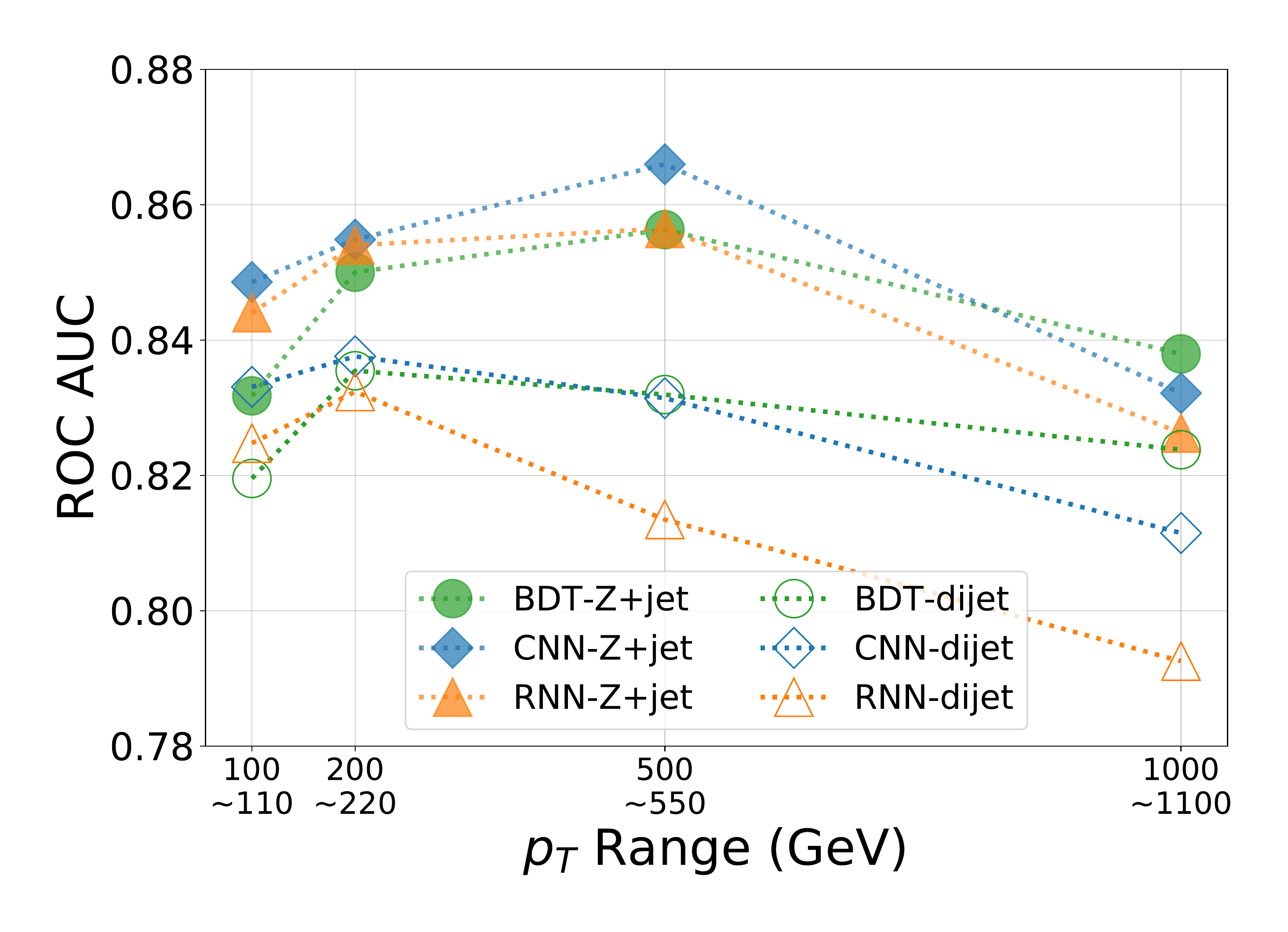}
\includegraphics[width=.4\linewidth]{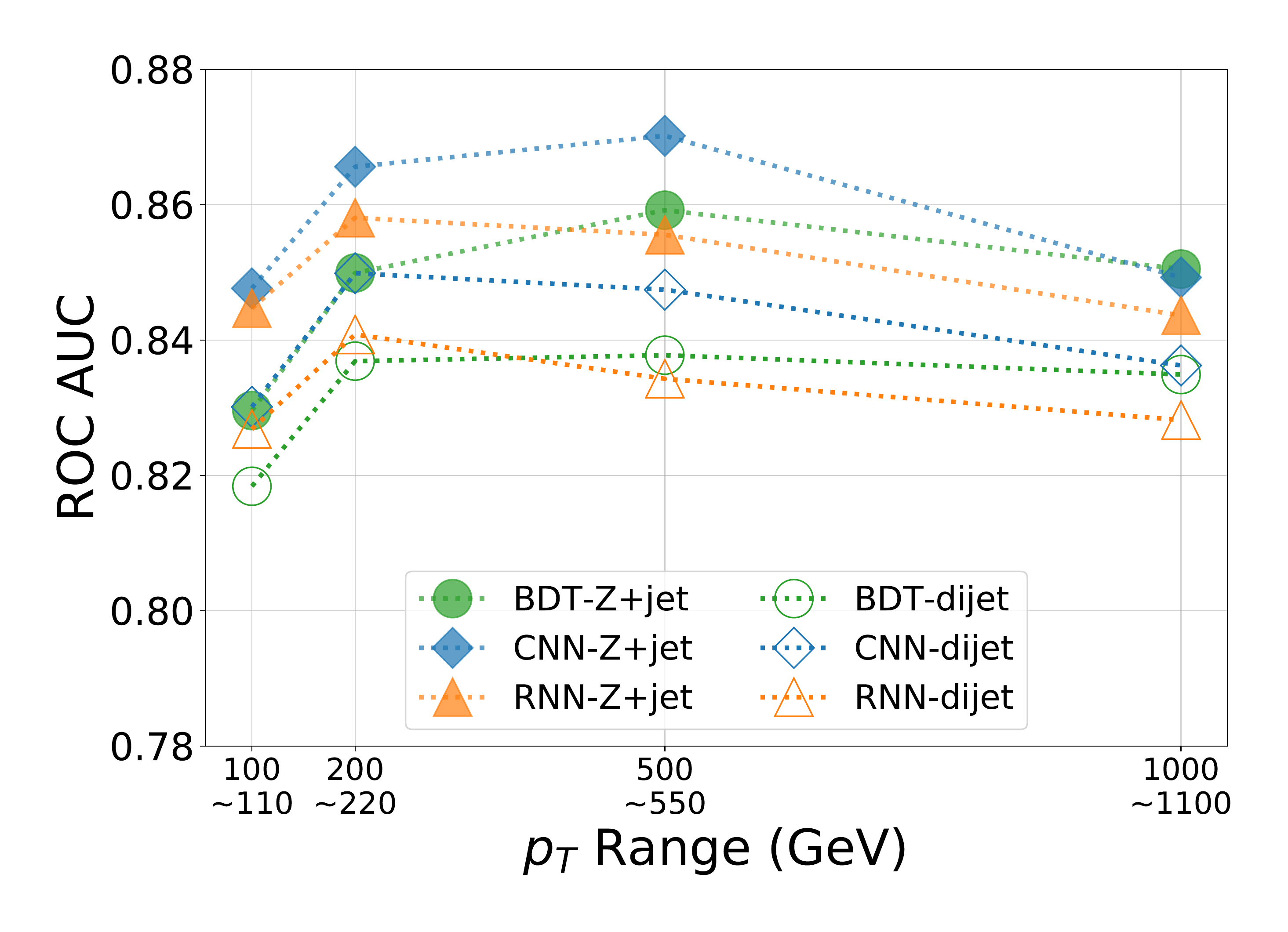}
\caption{ Area under the ROC curve (AUC) for the different pure quark or pure gluon testing samples and $p_T$ ranges. The training is done in all cases by using weakly supervised training on realistic samples and either using the full training sample (left) or a sample made requiring jets to pass a pseudorapidity cut (right). In both cases, the full test set (without pseudorapidity cut) is used to find the AUC.}
    \label{fig:pure}
\end{figure}

In order to improve the rather poor performance of the deep learning classifiers with weakly supervised learning, we also investigated other differences in the Z+jet and the dijet samples.
In particular, a difference is seen in the distribution of the $\eta$ of the jets, as is shown in Fig.~\ref{fig:etadist}.
Previously, we saw that this did not affect either the BDT variable distributions or the performance of the output of the fully supervised pure sample.
However, when we trained the deep learning models with the weakly supervised method and an additional selection of $|\eta| < 1$, the performance was greatly improved.
Figure~\ref{fig:pure} compares the AUC before and after the $\eta$ selection and shows both that the performance improves above that of the BDT and that the performance difference between the Z+Jet and the dijet samples reduces to the level of the fully supervised algorithm.
The deep learning models, therefore, appear to have recovered enough information to estimate the $\eta$ of the jet and can be used to further discriminate between the Z+jet and the dijet samples, beyond simply the quark-gluon jet fraction.
This shows the importance of cross-checks when using the weakly supervised method on data, as any difference in the sample beyond the quark-gluon jet fraction could degrade performance.

\section{Conclusions}

The performance of the CWoLa method with CNN, RNN and BDT models using realistic MC simulations was studied for discriminating between quark-initiated and gluon-initiated jets.
The realistic MC samples that were used were gluon-enriched dijet samples and quark-enriched Z+jet samples in several different jet $p_T$ ranges.
As expected, the classifiers trained to distinguish the Z+jet and the dijet sample were good quark-gluon jet classifiers, as shown for samples of pure quark or pure gluon jets.
However, unlike in previous studies which simply mixed quarks and gluons from a single source, we found that the simulated Z+jet and dijet samples had slightly different quark and gluon jet fragmentation properties.
These differences led to slightly decreased quark-gluon discrimination performance of the classifiers, even pure classifiers, on the dijet sample.

The weakly supervised deep learning classifiers also showed lower overall performance with increasing jet $p_T$ whereas classifiers trained with pure quark-gluon samples showed increased performance.
This is partly expected as the quark fraction of the dijet sample increases above 50\% at very high $p_T$ and the weakly supervised methods become harder to train as the fractional difference between the samples becomes smaller.
However, the performance of the BDT remained relatively good up to high $p_T$.
We investigated and found that when we restricted the $\eta$ range of the training to only central jets, we could recover the deep learning performance at high $p_T$, even when testing across the full $\eta$ range.
As the samples have different $\eta$ distributions, this is due to the deep learning methods being able to discover the $\eta$ of the jet based on the daughter particle information and to use that to further discriminate between the dijet and the Z+jet samples, rather than looking at just the quark-gluon differences.

We conclude that, based on realistically simulated Z+jet and dijet samples, weakly supervised learning could be applied to train a quark-gluon classifier using only data.
Several additional factors, such as additional particles from pile-up events entering the inputs and uncertainties due to the jet energy scale, could affect the overall conclusions and would need consideration with these methods.
Deep learning methods in particular, however, even at the level of this study, need to be carefully trained to avoid training the classifiers on differences other than the quark-gluon jet fraction between the samples.


\begin{acknowledgments}
This article was supported by the computing resources of the Global Science Experimental Data Hub Center (GSDC) at the Korea Institute of Science and Technology Information (KISTI). 
J.L. and I.W. are supported by the Korea Research Fellowship Program through the National Research Foundation of Korea (NRF) funded by the Ministry of Science and ICT (KRF project grant number: 2017H1D3A1A01052807).
I.P., Y.L. and S.Y. are supported by the Basic Science Research Program through the NRF funded by the Ministry of Education (2018R1A6A1A06024977).
\end{acknowledgments}


\end{document}